\documentclass{amsart}

\theoremstyle{definition}

\theoremstyle{remark}

\newcommand{\cC}{{\mathcal C}}
\newcommand{\cD}{{\mathcal D}}

\newcommand{\bR}{{\mathbb{R}}}

\numberwithin{equation}{section}

\begin{document}

\title{{On Quantum Lyapunov Exponents\footnote{\uppercase{S}upported by 
\uppercase{KBN grant PB/1490/PO3/2003/25 and RTN HPRN-CT}-2002-00279.}}}
\author{W{\l}adys{\l}aw A. Majewski}
\address{Institute of Theoretical Physics and Astrophysics\\ Gda{\'n}sk
University\\ Wita Stwo\-sza~57\\ 80-952 Gda{\'n}sk, Poland}
\email{fizwam@univ.gda.pl}
\author{Marcin Marciniak}
\email{matmm@univ.gda.pl}

\begin{abstract}
It was shown that quantum analysis constitutes the proper analytic basis for quantization of Lyapunov exponents in the Heisenberg picture. Differences among various quantizations of Lyapunov exponents are clarified.
\end{abstract}

\maketitle

\section{Introduction}
One of the most useful concepts in the theory of (classical) dynamical systems is the idea of Lyapunov exponent 
(see \cite{A,Eck,Haake,Gutz,P}).
It measures the average rate of growth of the separation of orbits which differ by a small vector at time zero.

For the non-commutative setting, this concept has many various generalizations which are called quantum Lyapunov exponents, QLE for short (see \cite{MK,ENTS,J,Fal,Majew,MV,T,V,V1} and references therein).
As there are different definitions of QLE it is of interest to know whether there exists
a common analytical basis for these generalizations. In particular, such basis could be very useful in establishing  relations among various quantizations schemes.

In this paper we intend to argue that so called Quantum Analysis (see \cite{S1,S2,S3,S4}) 
can be taken as such framework for the definitions which are given in the Heisenberg picture. Subsequently, we review several representative definitions and clarify differences among them.

\section{Preliminaries }

Let us set up notation and terminology.
The triple
\begin{equation}
(X,\tau: X \to X, \mu) \qquad  \Bigl[ (X, \tau_t:X \to X, \mu) \Bigr]
\end{equation}
defines discrete (continuous)  classical dynamical system where $X$ is a measurable space, $\mu$ a measure, and finally
$\tau$ ($\tau_t$) a measure preserving map (maps respectively). If $X$ is equipped with the differential calculus then the classical Lyapunov exponent is defined as
\begin{equation}
\label{Lapunov}
\lambda(x,y) = \lim_{n \to \infty} \frac{1}{n} \log|D_y \tau^n(x)| \qquad \Bigl[ 
\lambda(x,y) = \lim_{t \to \infty} \frac{1}{t} \log|D_y \tau_t(x)| \Bigr]
\end{equation}
where $D_y \tau^n(x)$ ($D_y \tau_t(x)$) denotes directional derivative of $\tau$ composed
with itself $n$ times (of $\tau_t$ respectively) at $x$.
By a non-commutative discrete (continuous) dynamical system we mean
\begin{equation}
({\frak A}, \tau: {\frak A} \to {\frak A}, \phi) \qquad \Bigl[ ({\frak A}, \tau_t: {\frak A} \to {\frak A}, \phi) \Bigr]
\end{equation}
where $\frak A$ is a $C^*$-algebra (with unit), $\tau$ ($\tau_t$) is a smooth enough positive map (maps respectively), and $\phi$ is a state. Let us note, that in general, $\tau$ ($\tau_t$) does not need to be a linear map.

Non-commutative dynamical systems, so in particular the concept of QLE, can be studied by means of quantum analysis, i.e. 
one can employ the analysis {\em using 
a non-commutative calculus of operator derivatives and integrals, where derivatives are defined within Banach space technique
on the basis of the Leibnitz rule, irrespective of their explicit representations such as the G\^ateaux derivative or commutator}, see
\cite{S1}, \cite{S3}, \cite{S4}.

In particular, putting $\delta_A \equiv [A, \cdot ]$, one can
verify (see \cite{S3}) that $\delta_{A \to B} \equiv - \delta_A^{-1} \delta_B$ is the well defined map 
satisfying the Leibnitz rule, when its domain
${\cD}_A$  consists of convergent power series of the operator $A$ (convergent in norm). Moreover, one can obtain the nice formula for a derivative $D$ (cf. \cite{S3}); namely
\begin{equation}
D(A\tau(A)) = D(\tau(A) A)
\end{equation}
when $\tau(A) \in {\cD}_A$. Then, one has
\begin{equation}
\delta_A D\tau(A) = \delta_{\tau(A)} DA = - \delta_{DA} \tau(A),
\end{equation}
hence
\begin{equation}
D\tau(A) = - \delta_A^{-1} \delta_{DA} \tau(A) \equiv \delta_{A \to DA} \tau(A).
\end{equation}

Finally, following Suzuki (see \cite{S3} for details) we define another kind of differential $d_{A \to B}$, satisfying
the Leibnitz rule for $B=dA$, with use of the partial inner derivation, i.e.
\begin{equation}
d_{A \to B} \equiv \delta_{(-\delta^{-1}_A B);A}
\end{equation}
and the commutator $\delta_{(- \delta^{-1}_A B)}$ is taken only with the operator $A$ in a multivariate operator $f(A,B)$. The domain $\cD_{A,B}$ of $d_{A \to B}$ is given by the set of convergent non commuting
power series of operators $A$ and $B$.

With these preparations, we will discuss various quantizations in which the concept of Lyapunov exponent can be introduced.
\section{Quantum Lyapunov exponents}
\subsection{A}
We begin with the first algebraic (and in fact very straightforward) generalization of Lyapunov exponent. It is defined here for discrete quantum dynamical systems only
(see \cite{MK}):
\begin{equation}
\label{MK}
\lambda^q(\tau, A, B) = \lim_{n \to \infty} \frac{1}{n} \log ||(D_B \tau^n)(A)||
\end{equation}
where we have used the G\^ateaux derivation.
Having defined QLE, $\lambda^q$, one should ask about its existence. For the quantization (\ref{MK}) this question splits naturally into two cases:

i) $\tau(\cdot)$ is a (non-linear) completely positive map.

ii) $\tau(\cdot)$ is a smooth positive (but not completely positive) function of $A$.

The case (i) was treated in \cite{MK}. We only remark that when $\tau$ is the sum of multilinear maps the product structure of derivatives is not so essential as for plain positive maps and the analysis of $\lambda^q$ can be done without using the quantum analysis.
On the contrary, in the case (ii) the analysis of existence of $\lambda^q$ needs the use of quantum analysis and as far as we know every model has been treated separately (see \cite{M}).

\subsection{B} As the second quantization we consider that which was given in \cite{ENTS}. Denote by $(\frak M, \tau_t)$ a quantum dynamical system based on von Neumann algebra $\frak M$ and on a distinguished evolution $\tau_t$.
Namely, let $\delta_j$ be the derivation generating the ``horocyclic'' action $\sigma_s^j$ on a von Neumann algebra $\frak M$, i.e.
for the map $ \bR \ni s \to \sigma_s^j \in Aut(\frak M)$ one has 
\begin{equation}
\label{horocyclic}
\tau_t \circ \sigma^j_s \circ \tau_{- t} = \sigma^j_{e^{- \lambda_j t} s}
\end{equation}
where $\lambda_j, s, t \in \bR$.

Then

\begin{equation}
\label{ENTS}
\lambda^E(j, A) = \lim_{t \to \infty} \frac{1}{t} \log ||\delta_j \tau_t(A)||
\end{equation}
can be considered as another quantization of the Lyapunov exponent. We note that the proof of existence of (\ref{ENTS}) is an easy task.
The discussion of other properties of $\lambda^E$ will be postponed to the final subsection and here we only note that the property (\ref{horocyclic})
is essential for the proper interpretation of this quantization.

\subsection{C}
The last quantization which we wish to consider was proposed by Jauslin et al \cite{J}.  To define it, let $\delta_{L_{\vec{\alpha}}}$
be the derivation generating the action $\sigma$ (induced by translations on $\bR^2$) on the Weyl algebra $\frak W$ constructed over $\bR^2$, i.e. $\bR \ni s \to \sigma_s \in {\mathrm{Aut}}({\frak W})$ is fixed. Consider the dynamical system $({\frak W}, \tau_t(A) \equiv U^*_t A U_t)$,
i.e.  the dynamics $\tau_t$ is implemented by the one parameter family of unitary operators $\{ U_t \}$. The number
\begin{equation}
\label{J}
\overline{\lambda}(\tau, L, A) = \sup_{\vec{\alpha} \in \bR^2} \limsup_{t \to \infty} \frac{1}{t} \log||\delta_{L_{\vec{\alpha}}}(\tau_t(A))||
\end{equation}
 was called the upper quantum Lyapunov exponent \cite{J}. Let us note that $\overline{\lambda}$ given by (\ref{J}) is defined for very particular dynamical model as well as for the specific action. Moreover, its existence is not guaranteed.
 
 \subsection{D}
Now we are in position to compare the above definitions as well as to elucidate the role of derivatives $\delta^j$ and $\delta_{L_{\vec \alpha}}$. To this end we will use the framework of Quantum Analysis in which the rules do not depend on explicit representation of derivations (cf. Section 2).

Let us recall the basic idea of Lyapunov exponents: we should study the evolution of a slight change of initial conditions for the considered dynamical map. 
Implementing this idea rigorously, one can study the variation of initial conditions resulting from implementation of an action 
 $\sigma
\in {\mathrm{Aut}} (\frak{M})$ (or in ${\mathrm{Aut}}(\frak W$)). Then, using the quantum analysis (see \cite{S3}) one is lead to investigate 
\begin{equation}
\frac{d}{ds} \tau_t(\sigma_s(A)) = \frac{d \tau_t(\sigma_s(A))}{d \sigma_s(A)} \frac{d \sigma_s(A)}{ds}
= \frac{d \tau_t(\sigma_s(A))}{d \sigma_s(A)} \delta(\sigma_s(A))
\end{equation}
where $\sigma$ denotes the action generated by the derivation $\delta$. But, this does not offer any significant elucidation of the role of $\delta_j$ and $\delta_{L_{\vec{\alpha}}}$. This provides a clarification for $D$ only.

However, the separation of initial conditions caused by the action $\sigma_s$, for small $s$ can also be analysed using the operator Taylor expansion
(cf. \cite{S3})

\begin{equation}
\tau_t(A + s \delta(A)) = \sum_{n=0}^{\infty} \frac{s^n}{n!} d^n_{A \to \delta(A)} \tau_t(A)
\end{equation}
Hence
\begin{equation}
\lim_{s \to 0} \frac{1}{s} \bigl( \tau_t(A + s\delta(A)) - \tau_t(A) \bigr) = d_{A \to \delta(A)} \tau_t(A) = \delta_{A \to \delta(A)} \tau_t(A)
\end{equation}
where the last equality follows from the formulas (2.26) in \cite{S3}.
Hence, quantizating along the lines given by (\ref{Lapunov}) one arrives at
\begin{equation}
\label{def1}
\lim_{t \to \infty} \frac{1}{t} \log ||\delta_{A \to \delta(A)} \tau_t(A) ||
\end{equation}
or
$$
\limsup_{t \to \infty} \frac{1}{t} \log ||\delta_{A \to \delta(A)} \tau_t(A) ||.$$
We stress again: the change of initial conditions is implemented by the action $\sigma_s$. Obviously, (\ref{def1}) can be considered as a particular case of definition (\ref{MK}).

Next, let us consider the case when the action $\sigma_s$ is applied to the orbit $\bR \ni t \mapsto \tau_t(A) \in \frak{A}$ for some $C^*$-algebra $\frak A$. Then, one has
\begin{equation}
\frac{d}{ds} \sigma_s \tau_t(A) |_{s=0} = \delta \tau_t(A)
\end{equation}

This leads to the following expression
\begin{equation}
\label{def2}
\limsup_{t \to \infty} \frac{1}{t} \log ||\delta \tau_t(A)||
\end{equation}
Clearly, (\ref{def2}) is the essential ingredient of definition (\ref{J}) given in subsection 3.3.
However, taking into account the way in which (\ref{def2}) was originated, {\em one could say that (\ref{J}) can be interpreted as a measure of asymptotic stability of the action $\sigma_s$ along the fixed trajectory $\bR \ni t \mapsto \tau_t(A) \in \frak{A}$ and not necessarily as Lyapunov exponent for $\tau_t$ in the strict sense.}

Turning to definition given by Emch et al (\ref{ENTS}), it should be noted that it is exactly the ``horocyclic'' property, $\tau_t \circ \sigma_s^j \circ \tau_{-t} = \sigma^j_{e^{-\lambda_jt}s}$ which enables us to treat this quantization along the lines of (\ref{def1}) - one can intertwine (``commute'') the evolution $\tau$ with the action $\sigma$.

We want to close this paper with a comment why definition (\ref{MK}) also can be considered as a quantization of the upper Lyapunov exponent (cf. \cite{MK2}).
To this end, let us pick a self-adjoint $A
\in {\frak A}$ where for simplicity we will assume that $\frak A$ is a concrete $C^*$-algebra, i.e. ${\frak A}
\subset B(H)$ for some Hilbert space $H$. 
${\frak A}_A$ will denote the $C^* $-algebra generated by $A$ and $\bf 1$. 
Obviously, ${\frak A}_A$ plays the role of $\cD_A$ (cf. Section 2). Let us
assume that $\tau ({\frak A}_A) \subset {\frak A}_A$  and that 
the restriction of ${\tau}^n$  to ${\frak A}_A$, 
${\tau}^n{\vert}_{{\frak A}_A}$,  has the Taylor expansion. Then, we can
write 
\begin{equation}
\tau (A) = \int_{{\sigma}(A)} {\tau}({\lambda}) dE({\lambda})
\end{equation}
where the spectral resolution of $A$ $(= \int_{{\sigma}(A)} {\lambda} dE({\lambda}))$ 
is used. Moreover,
\begin{equation} 
{\tau}^n (A) = \int_{{\sigma}(A)} {\tau}^{n}({\lambda}) dE({\lambda})
\end{equation}
and
\begin{equation}
\Vert D_{B}{\tau}^n (A)\Vert = \Vert \int_{{\sigma}(A)} D{\tau}^{n}({\lambda})b({\lambda}) dE({\lambda})\Vert 
\end{equation}
where $B \in {\frak A}_A$
 and the function $b({\lambda})$ is the image of $B$ with respect 
 to the ($^*$-spectral) isomorphism ${\phi} : {\frak A}_A \to {\cC}({\sigma}(A))$ with  
${\sigma}(A)$ denoting the spectrum of $A$. Let us    
 restrict ourselves to $B = {\bf 1}$. Then, one has
\begin{eqnarray}
\lefteqn{\lambda^q(\tau ; A, {\mathbf{1}}) =
\lim_{n\to\infty} {1\over
n} \log\Vert\int_{{\sigma}(A)} D{\tau}^{n}({\lambda})
dE({\lambda})\Vert }
               \nonumber\\
& & = \lim_{n\to\infty} {1\over
n} \sup_{{\lambda} \in {\sigma}(A)}\log\vert D{\tau}^{n}({\lambda})
\vert 
 = \lim_{n\to\infty} \sup_{{\lambda}}\,{1\over
n} \log\vert D{\tau}^{n}({\lambda})
\vert 
\end{eqnarray}
On the other hand, if
\begin{equation}
\label{9} 
\lambda^{\rm cl}({\tau}, {\lambda}) = \lim\limits_{n\to\infty}\,{1\over n}\,\log \vert D{\tau}^n ({\lambda}) \vert , 
\end{equation}
exists and the limit in (\ref{9}) is uniform with respect to $\lambda$ then
\begin{equation}
 \lambda^q(\tau ; A, \bf{1}) = \sup_{{\lambda}} \lambda^{\rm
cl}({\tau}, {\lambda}) 
\end{equation}
Therefore, we conclude that the norm used in definition (\ref{MK})
gives the quantum generalization of the largest
characteristic exponent. This legitimizes the claim that 
$\lambda^q(\tau, A, \bf{ 1})$ can also be considered as the quantization of the upper Lyapunov exponent.

\end{document}